# AN EFFICIENT DATA IMPUTATION TECHNIQUE FOR HUMAN ACTIVITY RECOGNITION

Ivan Miguel Pires
*Instituto de Telecomunicações*
*Universidade da Beira Interior*
*Covilhã, Portugal*
*Department of Computer Science*
*Polytechnic Institute of Viseu*
*Viseu, Portugal*
*impires@it.ubi.pt*

Faisal Hussain
*Department of Computer Engineering*
*University of Engineering and Technology (UET)*
*Taxila, Pakistan*
*faisal.hussain.engr@gmail.com*

Nuno M. Garcia
*Instituto de Telecomunicações*
*Universidade da Beira Interior*
*Covilhã, Portugal*
*ngarcia@di.ubi.pt*

Eftim Zdravevski
*Faculty of Computer Science and Engineering*
*University Ss Cyril and Methodius*
*Skopje, North Macedonia*
*eftim.zdravevski@finki.ukim.mk*

**ABSTRACT**

The tremendous applications of human activity recognition are surging its span from health monitoring systems to virtual reality applications. Thus, the automatic recognition of daily life activities has become significant for numerous applications. In recent years, many datasets have been proposed to train the machine learning models for efficient monitoring and recognition of human daily living activities. However, the performance of machine learning models in activity recognition is crucially affected when there are incomplete activities in a dataset, i.e., having missing samples in dataset captures. Therefore, in this work, we propose a methodology for extrapolating the missing samples of a dataset to better recognize the human daily living activities. The proposed method efficiently pre-processes the data captures and utilizes the k-Nearest Neighbors (KNN) imputation technique to extrapolate the missing samples in dataset captures. The proposed methodology elegantly extrapolated a similar pattern of activities as they were in the real dataset.

## 1. INTRODUCTION

The recent advancements of the technology (Cardinale and Varley, 2017; Kim et al., 2016; Ni Scanaill et al., 2011; Patel et al., 2012; Sousa et al., 2015) and the presence of sensors in the commonly used off-the-shelf mobile devices (Shahriyar et al., 2010; Stankevich et al., 2012; Steele, 2011; Tian et al., 2019; Ventola, 2014) allow the development of solutions for the identification of the human daily living activities in order to monitor its lifestyles, *e.g.*, the creation of a Personal Digital Life Coach (Garcia, 2016). However, the accuracy of these systems and their resilience of fails is essential for the recognition of activities in different environments (Dimitrievski et al., 2016a; Pires et al., 2017, 2018a; Zdravevski et al., 2015). In general, sportspeople, older adults, and other persons with special needs are living in conditions with bad network connection, but the development of solutions for this type of people is vital to improve their quality of life (Dimitrievski et al., 2016b; Sendra et al., 2012; Seneviratne et al., 2017).

There are different types of daily activities, and most of these are detected with inertial sensors, e.g., walking, running, moving upstairs, moving downstairs, and standing (Ferreira et al., 2020; Pires et al., 2018b, 2018c, 2018d, 2019, 2020). These are simple activities with different types of motion that can be distinguished with low-cost sensors. The combination of artificial intelligence methods and the data acquired sensors available in the off-the-shelf mobile devices may empower the monitoring of older adults, and the development of intelligent solutions for sports and medicine (Costa et al., 2015; E. Zdravevski et al., 2017; Kumar and Venkatesan, 2014).

Currently, the devices used for the data acquisition may fail due to the memory, battery, and power processing constraints (Pires et al., 2017, 2018a) which cause missing samples/values while data acquisition. Thus, the activities are wrongly recognized. However, the correct recognition of activities is essential to develop solutions that support various types of people in different daily activities. The implementation of data imputation minimizes these constraints measuring the data in faults. As the proposal uses three sensors (*i.e.,* accelerometer, magnetometer, and gyroscope), they have the same number of records and the same frequency of data acquisition to be fused and comparable. Before this study, we used different statistical features for the identification of activities and environments. Still, the data imputation allows the implementation of different techniques for the recognition of the activities, which may improve the accuracy of the recognition performed by our framework (Ferreira et al., 2020; Pires et al., 2018b, 2018d, 2019, 2020).

This paper's primary motivation is to increase the reliability of the previously proposed framework for the recognition of daily activities (Ferreira et al., 2020; Pires et al., 2020, 2019, 2018b, 2018d) with the imputation of the missing data to fill the data and correctly identify the daily activities.

This paper proposes implementing the k-Nearest Neighbors (KNN) imputation algorithm for the estimation of the values of the different datasets to fulfill the number of outputs correctly. The inertial sensors, including accelerometer, magnetometer, and gyroscope sensors, have three axes (*i.e.,* x, y, and z), where the measurement should be related to the different axis. The proposed method first identifies the missing data, then performs data segmentation, and, finally, imputes the missing data.

In the literature, the KNN imputation is one of the most used methods to support the recognition of activities and environments. With our implementation, it is verified that the pattern of the imputed data is similar but with higher amplitude than the original data.

This paper is organized as follows: Section 2 presents the methodology, including the definition of data acquisition, identification of missing samples, and implementation of data segmentation and imputation algorithms. The discussion and results are presented in Section 3. Finally, Section 4 presents the conclusions of the study.

## 2. METHODOLOGY

Figure 1 shows the flow diagram of the proposed methodology to extrapolate the missing samples in a human activity recognition dataset. The proposed method consists of four major stages, *i.e.,* data acquisition (section 2.1), missing samples identification (section 2.2), data segmentation (section 2.3), and data imputation (section 2.4). These stages are explained in the subsequent sections.

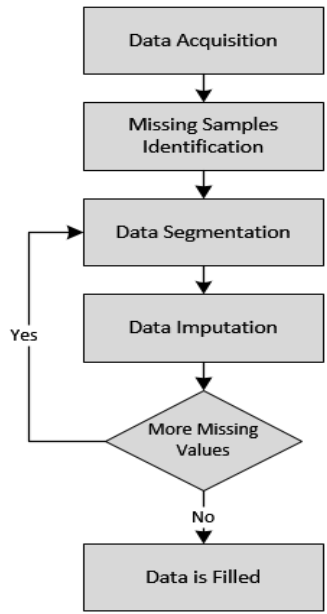

Figure 1. The workflow of the data imputation scheme

## 2.1. Data Acquisition

Data acquisition is the premier stage for performing the proceeding steps. In this stage, we first acquire the dataset for which we have to extrapolate the missing samples. Thus, we used a publicly available dataset (Pires, 2018). The dataset (Pires, 2018) includes five daily living activities, i.e., walking, running, standing, moving upstairs, and moving downstairs. The dataset (Pires, 2018) is acquired using three motion sensors, *i.e.,* accelerometer, gyroscope, and magnetometer. The five daily living activities included in the dataset (Pires, 2018) are performed by 25 subjects age ranging from (20-60) with sedentary and active lifestyles.

This dataset contains several captures with 5 seconds of sensors' data acquired every 5 minutes. While having a smartphone in the pocket, each activity was recorded by using an Android application. The data was obtained by hundreds of hours to implement methods for the identification of the activities. Due to the memory, battery, and power processing constraints, sometimes the data acquisition may fail (Pires et al., 2017, 2018a), and the imputation of the mission data is needed.

## 2.2. Missing Samples Identification

Once the dataset (Pires, 2018) is acquired, the next step is to check if there are some missing samples in each recorded activity. The missing samples in a dataset have a lousy impact while training the machine learning (ML) models. It occurs because due to missing samples, the ML models do not learn the activity patterns properly. The missing samples in the dataset are mainly due to two reasons, *i.e.,* the user did not perform an activity for a complete defined activity duration, or there could have some issue in the device that is being used for recording the activity.

To identify the missing samples in the dataset, we first analyzed each activity time duration and frequency rate. In our case, each activity was performed for a time duration of 5 seconds. The frequency rate for accelerometer and gyroscope was 100 Hz (*i.e.,* 100 samples/s), while for magnetometer, it was 10 Hz (*i.e.,* 10 samples/s). So, for a 5sec activity, there should be 5 x 100 = 500 samples for the accelerometer. On the other hand, there should be 5 x 10 = 50 samples for the gyroscope in an activity of 5 seconds.

Once we figured out the number of samples for each activity across the three sensors, we then write a python script to identify the missing samples in each activity to perform further steps. While recognizing the missing samples in an activity, we ignored the activities having missing samples for duration more than 1sec, i.e., ignored the captures that have more than 100 missing samples in case of accelerometer or more than 10 missing samples in case of gyroscope. It is done to be closer to the originality of the data than filling all synthetic samples to fulfill the space of missing samples.

**2.3. Data Segmentation**

After identifying the missing samples, we first inserted Null values to fulfill the space of missing samples. Then, we segmented the samples based upon the missing samples count. If the missing samples count was greater than 10, we segmented data into a chunk of 100 samples to include 90 original samples and the first 10 samples with Null values. If the missing samples count is less than or equal to 10 samples, we made a chunk of the last 100 samples, including the recently inserted and the Null value samples and original samples.

**2.4. Data Imputation**

Once the data is segmented, we then applied the KNN imputation algorithm (Beretta and Santaniello, 2016) to extrapolate the missing values. The KNN impuation technique is based upon the KNN algorithm. In KNN imputation, we firstly find k-closest neighbors to the missing data and then impute these missing values based upon known k-closest neighbors.

We noticed the count of missing samples every time, before applying the KNN imputation. If the missing samples were less than 10 samples, then the missing data is filled in the first iteration. However, if the missing samples are more than 10, we need to perform segmentation steps again to make another chunk of data and apply the KNN imputer algorithm still to extrapolate the missing values. As shown in Figure 1, this process continues until all the missing samples are extrapolated.

## 3. RESULTS AND DISCUSSION

The proposed data imputation technique is applied to a publicly available dataset (Pires, 2018) of human activity recognition. Based on the time of 5 seconds (500 samples per file in case of accelerometer and gyroscope and 50 samples per file in case of magnetometer), the dataset had many missing samples. Table 1 shows the missing samples in dataset (Pires, 2018) for the accelerometer data across each activity. In Table 1, the second column shows the total number of records available in the dataset (Pires, 2018) across each activity. The proceeding columns shows the number of missing samples in the dataset (Pires, 2018) for each corresponding activity. It can be verified that the most missing samples are related to moving upstairs activity.

| Activity | Total Records | Records having missing samples | | |
|---|---|---|---|---|
| | | <=10 | >10 & <=100 | >100 |
| Walking | 783 | 317 | 327 | 139 |
| Running | 487 | 403 | 72 | 12 |
| Standing | 399 | 360 | 37 | 2 |
| Moving upstairs | 707 | 173 | 115 | 419 |
| Moving downstairs | 364 | 111 | 75 | 178 |

Table 1. Accelerometer Missing Samples Statistics in Dataset.

To cope with this challenge, we proposed an imputation technique for extrapolating the missing samples by following the steps as described in the methodology section. Figure 2 presents an excerpt of accelerometer data for moving downstairs activity, taken from the dataset. The selected excerpt had 50 missing samples. We first applied the proposed methodology to identify the missing samples to validate the actual missing count. The proposed method accurately identified the correct missing samples and found 50 missing samples in the given excerpt.

After discovering that 50 samples are missing in the given excerpt, we then filled the missing samples by NULL values, as shown in Figure 3, for further implementation of the KNN imputation algorithm. Afterwards, we performed data segmentation, as explained in Section 2.3. Figure 4 shows how the data is segmented for the given excerpt.

Finally, we applied the KNN imputation algorithm to extrapolate the unknown samples, as explained in the previous sections. Figure 5 shows the missing calculated values extrapolated after using the KNN imputation algorithm.

| Sr.# | Timestamp | Ax | Ay | Az |
|---|---|---|---|---|
| 1 | 1493996698893 | -2.145 | -9.174 | 3.802 |
| 2 | 1493996698902 | -0.612 | -9.625 | 3.984 |
| 3 | 1493996698914 | -0.641 | -10.678 | 3.84 |
| | - - - | | | |
| 448 | 1493996702662 | -0.641 | -8.533 | 3.84 |
| 449 | 1493996702663 | -0.632 | -8.399 | 3.61 |
| 450 | 1493996702672 | -0.526 | -8.322 | 3.39 |

Figure 2. Excerpt of accelerometer data for walking activity.

| 450 | 1493996702672 | -0.526 | -8.322 | 3.39 |
|---|---|---|---|---|
| 451 | 1493996702682 | Null | Null | Null |
| 452 | 1493996702692 | Null | Null | Null |
| | - - - | | | |
| 499 | 1493996703162 | Null | Null | Null |
| 500 | 1493996703172 | Null | Null | Null |

Figure 3. Missing samples filled as NULL values

| 441 | 1493996702582 | -0.201 | -8.169 | 3.352 |
|---|---|---|---|---|
| 442 | 1493996702593 | -0.392 | -8.102 | 3.208 |
| 443 | 1493996702604 | -0.507 | -8.207 | 3.112 |
| | - - - | | | |
| 449 | 1493996702663 | -0.632 | -8.399 | 3.61 |
| 450 | 1493996702672 | -0.526 | -8.322 | 3.39 |
| 451 | 1493996702682 | Null | Null | Null |
| 452 | 1493996702692 | Null | Null | Null |

Figure 4. Data segmentation before applying the KNN imputation technique

| 449 | 1493996702663 | -0.632 | -8.399 | 3.61 |
|---|---|---|---|---|
| 450 | 1493996702672 | -0.526 | -8.322 | 3.39 |
| 451 | 1493996702682 | -0.329 | -9.866 | -1.05 |
| 452 | 1493996702692 | -0.332 | -9.197 | 2.405 |

Figure 5. Missing samples filled after applying the KNN imputation technique.

This technique is implemented for the files that have less than 100 missing samples in case of accelerometer and gyroscope and 10 missing samples in case of magnetometer. If more than 100 samples are missed, this capture should be discarded. In short, we ignored the activities having missing samples for duration more than 1sec, i.e., ignored the captures that have more than 100 missing samples in case of accelerometer or more than 10 missing samples in case of gyroscope. It is done to be closer to the originality of the data than filling all synthetic samples to fulfill the space of missing samples as explained in (section 2.2).

Figure 6 shows an example of an activity recorded while moving downstairs, where it is possible to verify that the file has 375 samples. As the file must have 500 samples, 125 samples are missing. By applying KNN Imputation methods, we found data with the same format, but it has more high amplitude. Thus, the peaks for the downstairs activity are similar, as presented in Figure 7.

For the future work, the data classification should be implemented with different machine learning methods, where we can now use different variables of previous works, *e.g.,* the mean, standard deviation, variation, and median of the values acquired and imputed from each axis. Thus, we can discover different patterns to improve the development of a method for the recognition of daily activities and environments.

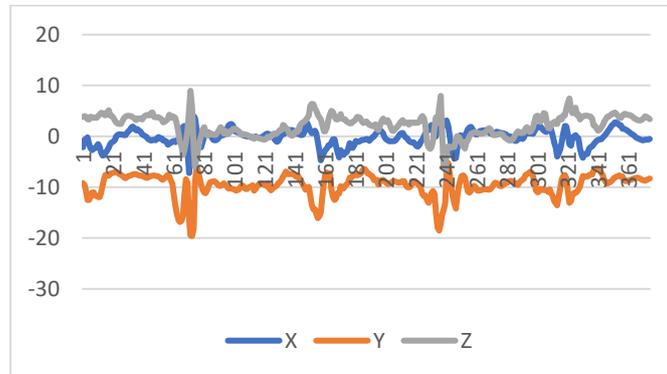

Figure 6. Accelerometer sample related to moving downstairs.

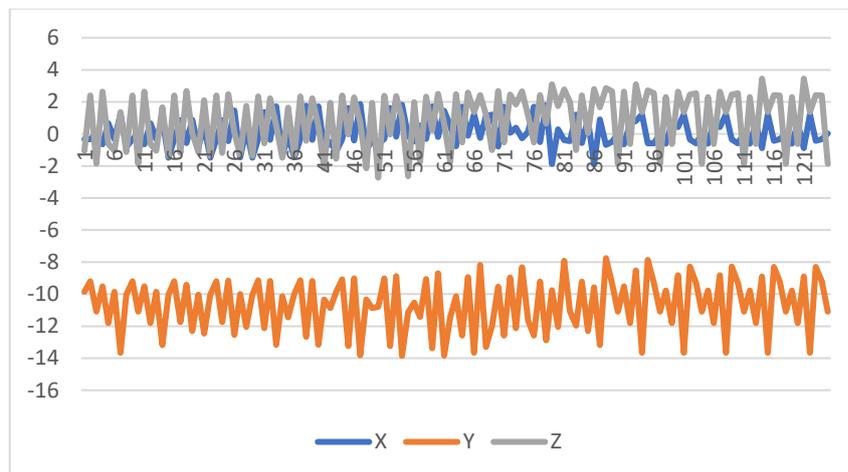

Figure 7. Imputed data for the accelerometer sample related to moving downstairs

## 4. CONCLUSION

The missing samples in a dataset capture affect the performance of machine learning models. Therefore, in this work, we proposed a methodology to extrapolate the missing samples of every dataset capture related to motion sensors data recorded for human activity recognition. The proposed method first identifies the missing samples in each excerpt with respect to time duration and frequency of the dataset. Secondly, it inserts Null values to fil the gap of missing samples. Thirdly, it performs the segmentation of the data, and finally, it applies the KNN imputation technique to extrapolate the missing samples. The proposed methodology extrapolated the missing samples with the same pattern as found in the original dataset but with some high amplitude.

As future work, we aim to improve the data amplitude to make it closer to the reality and train different machine learning models to better detect human activity patterns.

# ACKNOWLEDGEMENT

This work is funded by FCT/MEC through national funds and when applicable co-funded by FEDER-PT2020 partnership agreement under the project **UIDB/EEA/50008/2020**. (*Este trabalho é financiado pela FCT/MEC através de fundos nacionais e cofinanciado pelo FEDER, no âmbito do Acordo de Parceria PT2020 no âmbito do projeto UIDB/EEA/50008/2020*).

This article is based upon work from COST Action IC1303-AAPELE—Architectures, Algorithms and Protocols for Enhanced Living Environments and COST Action CA16226–SHELD-ON—Indoor living space improvement: Smart Habitat for the Elderly, supported by COST (European Cooperation in Science and Technology). COST is a funding agency for research and innovation networks. Our Actions help connect research initiatives across Europe and enable scientists to grow their ideas by sharing them with their peers. It boosts their research, career and innovation. More information in www.cost.eu.

# REFERENCES

Beretta, L., Santaniello, A., 2016. Nearest neighbor imputation algorithms: a critical evaluation. BMC Med Inform Decis Mak 16. https://doi.org/10.1186/s12911-016-0318-z

Cardinale, M., Varley, M.C., 2017. Wearable Training-Monitoring Technology: Applications, Challenges, and Opportunities. International Journal of Sports Physiology and Performance 12, S2-S2-62. https://doi.org/10.1123/ijspp.2016-0423

Costa, S.E.P., Rodrigues, J.J.P.C., Silva, B.M.C., Isento, J.N., Corchado, J.M., 2015. Integration of Wearable Solutions in AAL Environments with Mobility Support. J Med Syst 39, 184. https://doi.org/10.1007/s10916-015-0342-z

Dimitrievski, A., Zdravevski, E., Lameski, P., Trajkovik, V., 2016a. Towards application of non-invasive environmental sensors for risks and activity detection, in: 2016 IEEE 12th International Conference on Intelligent Computer Communication and Processing (ICCP). Presented at the 2016 IEEE 12th International Conference on Intelligent Computer Communication and Processing (ICCP), pp. 27–33. https://doi.org/10.1109/ICCP.2016.7737117

Dimitrievski, A., Zdravevski, E., Lameski, P., Trajkovik, V., 2016b. A survey of Ambient Assisted Living systems: Challenges and opportunities, in: 2016 IEEE 12th International Conference on Intelligent Computer Communication and Processing (ICCP). Presented at the 2016 IEEE 12th International Conference on Intelligent Computer Communication and Processing (ICCP), pp. 49–53. https://doi.org/10.1109/ICCP.2016.7737121

E. Zdravevski, P. Lameski, V. Trajkovik, A. Kulakov, I. Chorbev, R. Goleva, N. Pombo, N. Garcia, 2017. Improving Activity Recognition Accuracy in Ambient-Assisted Living Systems by Automated Feature Engineering. IEEE Access 5, 5262–5280. https://doi.org/10.1109/ACCESS.2017.2684913

Ferreira, J.M., Pires, I.M., Marques, G., García, N.M., Zdravevski, E., Lameski, P., Flórez-Revuelta, F., Spinsante, S., Xu, L., 2020. Activities of Daily Living and Environment Recognition Using Mobile Devices: A Comparative Study. Electronics 9, 180. https://doi.org/10.3390/electronics9010180

Garcia, N.M., 2016. A Roadmap to the Design of a Personal Digital Life Coach, in: Loshkovska, S., Koceski, S. (Eds.), ICT Innovations 2015, Advances in Intelligent Systems and Computing. Springer International Publishing, Cham, pp. 21–27. https://doi.org/10.1007/978-3-319-25733-4_3

Kim, S., Gajos, K.Z., Muller, M., Grosz, B.J., 2016. Acceptance of mobile technology by older adults: a preliminary study, in: Proceedings of the 18th International Conference on Human-Computer Interaction with Mobile Devices and Services. Presented at the MobileHCI '16: 18th International Conference on Human-Computer Interaction with Mobile Devices and Services, ACM, Florence Italy, pp. 147–157. https://doi.org/10.1145/2935334.2935380

Kumar, K.M., Venkatesan, R.S., 2014. A design approach to smart health monitoring using android mobile devices, in: 2014 IEEE International Conference on Advanced Communications, Control and Computing Technologies. Presented at the 2014 IEEE International Conference on Advanced Communications, Control and Computing Technologies, pp. 1740–1744. https://doi.org/10.1109/ICACCCT.2014.7019406

Ni Scanaill, C., Garattini, C., Greene, B.R., McGrath, M.J., 2011. Technology Innovation Enabling Falls Risk Assessment in a Community Setting. Ageing Int 36, 217–231. https://doi.org/10.1007/s12126-010-9087-7

Patel, S., Park, H., Bonato, P., Chan, L., Rodgers, M., 2012. A review of wearable sensors and systems with application in rehabilitation. Journal of NeuroEngineering and Rehabilitation 9, 21. https://doi.org/10.1186/1743-0003-9-21

Pires, I., Felizardo, V., Pombo, N., Garcia, N.M., 2017. Limitations of energy expenditure calculation based on a mobile phone accelerometer, in: 2017 International Conference on High Performance Computing & Simulation (HPCS). IEEE, pp. 124–127.

Pires, I.M., 2018. impires/August_2017-_Multi-sensor_data_fusion_in_mobile_devices_for_the_identification_of_activities_of_dail [WWW Document]. GitHub. URL https://github.com/impires/August_2017-_Multi-

sensor_data_fusion_in_mobile_devices_for_the_identification_of_activities_of_dail (accessed 5.8.20).

Pires, I.M., Garcia, N.M., Pombo, N., Flórez-Revuelta, F., 2018a. Limitations of the Use of Mobile Devices and Smart Environments for the Monitoring of Ageing People, in: HSP 2018.

Pires, I.M., Garcia, N.M., Pombo, N., Flórez-Revuelta, F., 2018b. Framework for the Recognition of Activities of Daily Living and their Environments in the Development of a Personal Digital Life Coach.

Pires, I.M., Garcia, N.M., Pombo, N., Flórez-Revuelta, F., Spinsante, S., Teixeira, M.C., 2018c. Identification of activities of daily living through data fusion on motion and magnetic sensors embedded on mobile devices. Pervasive and Mobile Computing 47, 78–93.

Pires, I.M., Marques, G., Garcia, N.M., Flórez-Revuelta, F., Canavarro Teixeira, M., Zdravevski, E., Spinsante, S., Coimbra, M., 2020. Pattern Recognition Techniques for the Identification of Activities of Daily Living Using a Mobile Device Accelerometer. Electronics 9, 509. https://doi.org/10.3390/electronics9030509

Pires, I.M., Marques, G., Garcia, N.M., Pombo, N., Flórez-Revuelta, F., Spinsante, S., Teixeira, M.C., Zdravevski, E., 2019. Recognition of Activities of Daily Living and Environments Using Acoustic Sensors Embedded on Mobile Devices. Electronics 8, 1499.

Pires, I.M., Teixeira, M.C., Pombo, N., Garcia, N.M., Flórez-Revuelta, F., Spinsante, S., Goleva, R., Zdravevski, E., others, 2018d. Android Library for Recognition of Activities of Daily Living: Implementation Considerations, Challenges, and Solutions.

Sendra, S., Granell, E., Lloret, J., Rodrigues, J.J.P.C., 2012. Smart collaborative system using the sensors of mobile devices for monitoring disabled and elderly people, in: 2012 IEEE International Conference on Communications (ICC). Presented at the 2012 IEEE International Conference on Communications (ICC), pp. 6479–6483. https://doi.org/10.1109/ICC.2012.6364935

Seneviratne, S., Hu, Y., Nguyen, T., Lan, G., Khalifa, S., Thilakarathna, K., Hassan, M., Seneviratne, A., 2017. A Survey of Wearable Devices and Challenges. IEEE Communications Surveys Tutorials 19, 2573–2620. https://doi.org/10.1109/COMST.2017.2731979

Shahriyar, R., Bari, Md.F., Kundu, G., Ahamed, S.I., Akbar, Md.M., 2010. Intelligent Mobile Health Monitoring System (IMHMS), in: Kostkova, P. (Ed.), Electronic Healthcare, Lecture Notes of the Institute for Computer Sciences, Social Informatics and Telecommunications Engineering. Springer Berlin Heidelberg, Berlin, Heidelberg, pp. 5–12. https://doi.org/10.1007/978-3-642-11745-9_2

Sousa, P.S., Sabugueiro, D., Felizardo, V., Couto, R., Pires, I., Garcia, N.M., 2015. mHealth Sensors and Applications for Personal Aid, in: Adibi, S. (Ed.), Mobile Health, Springer Series in Bio-/Neuroinformatics. Springer International Publishing, Cham, pp. 265–281. https://doi.org/10.1007/978-3-319-12817-7_12

Stankevich, E., Paramonov, I., Timofeev, I., 2012. Mobile phone sensors in health applications, in: 2012 12th Conference of Open Innovations Association (FRUCT). Presented at the 2012 12th Conference of Open Innovations Association (FRUCT) and Seminar on e-Travel, IEEE, Oulu, pp. 1–6. https://doi.org/10.23919/FRUCT.2012.8122097

Steele, R., 2011. Social media, mobile devices and sensors: Categorizing new techniques for health communication, in: 2011 Fifth International Conference on Sensing Technology. Presented at the 2011 Fifth International Conference on Sensing Technology, pp. 187–192. https://doi.org/10.1109/ICSensT.2011.6136960

Tian, S., Yang, W., Grange, J.M.L., Wang, P., Huang, W., Ye, Z., 2019. Smart healthcare: making medical care more intelligent. Global Health Journal 3, 62–65. https://doi.org/10.1016/j.glohj.2019.07.001

Ventola, C.L., 2014. Mobile Devices and Apps for Health Care Professionals: Uses and Benefits. P T 39, 356–364.

Zdravevski, E., Lameski, P., Kulakov, A., Kalajdziski, S., 2015. Transformation of nominal features into numeric in supervised multi-class problems based on the weight of evidence parameter, in: 2015 Federated Conference on Computer Science and Information Systems (FedCSIS). Presented at the 2015 Federated Conference on Computer Science and Information Systems (FedCSIS), pp. 169–179. https://doi.org/10.15439/2015F90